\documentstyle[12pt]{article}
\begin{document}
\pagestyle{empty}
\setlength{\oddsidemargin}{0.5cm}
\setlength{\evensidemargin}{0.5cm}
\setlength{\footskip}{1.5cm}
\renewcommand{\thepage}{-- \arabic{page} --}
\vspace*{-2.5cm}
\begin{flushright}
TOKUSHIMA 95-03 \\ (hep-ph/9505249) \\ May 1995
\end{flushright}
\vspace*{1cm}

\centerline{\large\bf Indications on the Higgs-Boson Mass from the
LEP Data}

\vspace*{1.5cm}

\begin{center}
\renewcommand{\thefootnote}{*)}
{\sc M. CONSOLI$^{\: a),\: }$}\footnote{E-mail address:
consoli@vxcern.cern.ch}
and
\renewcommand{\thefootnote}{**)}
{\sc Z. HIOKI$^{\: b),\: }$}\footnote{E-mail address:
hioki@ias.tokushima-u.ac.jp}
\end{center}

\vspace*{1.5cm}
\centerline{\sl $a)$ Istituto Nazionale di Fisica Nucleare - Sezione
di Catania}
\centerline{\sl Corso Italia, 57 - I 95129 Catania - ITALY}

\vskip 0.3cm
\centerline{\sl $b)$ Institute of Theoretical Physics,\ University of
Tokushima}
\centerline{\sl Tokushima 770 - JAPAN}

\vspace*{2.7cm}

\centerline{ABSTRACT}

\vspace*{0.4cm}
\baselineskip=20pt plus 0.1pt minus 0.1pt
We update our previous analysis on the Higgs mass $m_h$ and the QCD
coupling $\alpha_s(=\alpha_s(M_z))$ by using the LEP data after the
1995 Winter Conferences. For $m_t=180$ GeV we find evidence for a
rather large value of the Higgs mass in the range 500-1000 GeV, in
agreement with the indications from the W mass.

\vfill
\newpage
\pagestyle{plain}
\renewcommand{\thefootnote}{\sharp\arabic{footnote}}
\setcounter{footnote}{0}
\baselineskip=21.0pt plus 0.2pt minus 0.1pt

Strong evidence for the top quark has been observed by CDF and D0
Collaborations independently \cite{CDF}. We have now only one
yet-undiscovered particle left in the framework of the standard
electroweak model. A lot of experimental and theoretical efforts
should be made toward this particle, i.e., the Higgs boson. It is not
that easy to draw its indirect information from existing experimental
data since the Higgs mass $m_h$ enters the one-loop electroweak
predictions only logarithmically. Therefore, at present, one can only
hope to separate out the heavy Higgs-mass range (say $m_h\sim$
500-1000 GeV) from the low mass regime $m_h\sim$100 GeV as predicted,
for instance, from supersymmetric theories. Such analyses are,
however, still very important and indispensable for future
experiments at, e.g., LHC/NLC.

In our previous work \cite{zenro}, we have performed a detailed
comparison of the LEP data presented at the 1994 Glasgow Conference
\cite{oldlep} with the standard model for the various observables.
There we obtained some interesting information on $m_h$ and the
strong-interaction coupling constant at the Z-mass scale $\alpha_s=
\alpha_s(M_z)$. In this note, we shall update this analysis by using
the more precise data from the 1995 Winter Conferences as reported by
ALEPH, DELPHI, L3 and OPAL in \cite{LEP} and the new top mass $m_t=
180\pm 12$ GeV \cite{CDF}.

For our analysis, we used in \cite{zenro} the disaggregated data,
just as presented by the experimental Collaborations, without
attempting any average of the various results. This type of analysis
is interesting by itself to point out the indications of the various
sets of data since even a single measurement, if sufficiently
precise, can provide precious information. At the same time, since
the LEP data are becoming so precise, before attempting any averaging
procedure one should first analyze the various measurements with
their errors and check that the distribution of the results fulfills
the requirements of gaussian statistics. Without this preliminary
analysis one may include uncontrolled systematic effects which can
sizeably affect the global averages.

For instance, in \cite{ferroni} a detailed analysis of the relative
magnitude of the hadronic and leptonic widths for the different
channels of the various experiments was performed. Starting from the
LEP data presented at the Glasgow Conference \cite{oldlep} and using
a Monte Carlo method to generate a large number of ``a priori"
equivalent copies, one finds \cite{ferroni} that the probability of
the original LEP population is extremely small ($3.8\times 10^{-4}$).
Therefore, the meaning of the global average $R=\Gamma_h/\Gamma_l=
20.795\pm0.040$ presented in \cite{oldlep} is unclear and substantial
systematic effects have to be invoked to understand the distribution
of the various measurements.

In the following, we develop analyses similar way to our previous
work \cite{zenro} in order to make the comparison with it easy and
convenient. We shall first restrict to a fixed value of the top-quark
mass $m_t=180$ GeV and discuss the indications for the Higgs mass.
The experimental data relevant for our analysis are presented in
Table I. These are the available, individual results from the various
Collaborations as quoted in \cite{LEP} and the meaning of the various
quantities is the same as in \cite{LEP}. The theoretical predictions
in Table II, for several values of $\alpha_s$ and $m_h$
representative of the overall situation, have been obtained with the
computer code TOPAZ0 by Montagna et al. \cite{TOPAZ0}. Finally, in
Tables III-VI we report the partial and total $\chi^2$ for the
various experiments and in Table VII the sum of the $\chi^2$ for the
four Collaborations.

\vskip -0.15cm
\centerline{\bf ------------------------------}
\centerline{\bf Tables I -- VII}
\centerline{\bf ------------------------------}

We find here again some tendencies in the data: The global values of
the $\chi^2$ in Table VII confirm that $\alpha_s$ lies at $\sim
3\sigma$ from the DIS prediction $\alpha_s=0.113\pm 0.005$ (here, our
result is in very good agreement with the general analysis of
\cite{langa} which gives $\alpha_s=0.127\pm 0.005$). Further, by
inspection of Table I one finds evidences for some systematic effect
in the $\tau$ F-B asymmetry. This effect seems to be common to all
experiments and it is confirmed by the following remark. Let us
consider the global averages reported in \cite{LEP}
$$ A^{o}_{FB}(e)=0.0154\pm0.0030,                 \eqno(1) $$
$$ A^{o}_{FB}(\mu)=0.0160\pm0.0017,               \eqno(2) $$
$$ A^{o}_{FB}(\tau)=0.0209\pm0.0024,              \eqno(3) $$
and transform the averages for $A_e$ and $A_{\tau}$ \cite{LEP}
$$ A_e=0.137\pm0.009,~~~A_{\tau}=0.140\pm0.008    \eqno(4) $$
into ``equivalent" F-B asymmetries by using the standard model
formula
$$ A^{o}_{FB}(1)={{3}\over{4}}A^2_e,~~~
   A^{o}_{FB}(2)={{3}\over{4}}A_e A_{\tau}.       \eqno(5) $$
We find
$$ A^{o}_{FB}(1)=0.0141\pm0.0019,~~~
   A^{o}_{FB}(2)=0.0144\pm0.0018                  \eqno(6) $$
in very good agreement with Eqs.(1,2) but not with Eq.(3). Therefore,
there may be some problem in the direct measurement of
$A^{o}_{FB}(\tau)$ since all other measurements are in excellent
agreement with each other. Just to have an idea of the effect, if the
data for the $\tau$ F-B asymmetry are omitted in the evaluation of
the $\chi^2$ we find the results illustrated in Table VIII which one
should compare with Tables III- VII. The ``bulk" of the LEP data,
namely those well consistent with each other, show no preference for
a light Higgs boson and the best values of the $\chi^2$ are obtained
for a large value of $m_h$, just as in the case of the W mass
reported in \cite{hioki}.\footnote{The latest world average of the
   W-mass is $M_w=80.27\pm 0.14$ GeV. Comparing it with the one
   computed from $M_z$, we find not only that the central value of
   $m_h$ must be more than 1 TeV but also that $m_h=100$ GeV is
   disfavored (though at $1\sigma$ level). See \cite{hioki2} for
   more details.}

\vskip -0.15cm
\centerline{\bf ------------------------------}
\centerline{\bf Table VIII}
\centerline{\bf ------------------------------}

Finally, to have an idea of the dependence on $m_t$, we report in
Table IX and Table X the total $\chi^2$ for $m_t$=170, 180 and 190
GeV including all data or excluding $A^{o}_{FB}(\tau)$. As first
noticed by Ellis et al. \cite{ellis,fogli}, by increasing
(decreasing) the top-quark mass a larger (smaller) value of $m_h$ is
favoured and the shape of the $\chi^2$ is well consistent with all
values of the Higgs mass. For $m_t=180$ GeV, however, Table IX and
Table X give rather different information and it becomes crucial to
include the more problematic data for $A^{o}_{FB}(\tau)$ to
accommodate values $m_h\sim$ 100 GeV.

\vskip -0.15cm
\centerline{\bf ------------------------------}
\centerline{\bf Tables IX and X}
\centerline{\bf ------------------------------}

We have of course no mind to say that Tables VIII and X represent a
more faithful representation of the real physical situation than
Tables VII and IX. Most likely, our results suggest only that further
improvement in the data taking is needed for a definitive answer. We
may, however, conclude that it is dangerous to focus on a light-mass
region in Higgs searches at future experiments. Also, our analysis,
confirming the conclusions of \cite{zenro}, shows that the
possibility to obtain precious information on the Higgs mass is not
unrealistic when the top-quark mass will be measured with a higher
precision at the Tevatron.

To better appreciate this point, let us consider the hypothetical
situation where $m_t$ would be known to be 180 GeV to very high
accuracy (at the end of the century the combined CDF+D0 determination
should provide an overall error $\Delta m_t=\pm 3$ GeV \cite{mlm}).
In this case, what would we deduce from the present LEP results? On
one hand, we find a clear signal for a heavy Higgs from the very
precise data of the OPAL Collaboration (see Table VI) which
completely confirms the indications from the W mass. In fact, by
inspection of Table VI, the two pairs corresponding to $m_h=100$ GeV
lie outside the 95$\%$ C.L. contour ($\Delta\chi^2=+6.1$) in the
two-parameter space $(\alpha_s,m_h)$. This effect is independent on
$\alpha_s$ since the pair $(0.13, 100)$ has a total $\chi^2=13.87$
with a difference $\Delta\chi^2=+7$ with respect to the configuration
$(0.13,1000)$ shown in Table VI. On the other hand, in this
hypothetical scenario for the top mass, the OPAL trend is not
confirmed by the other Collaborations since, by subtracting out the
OPAL data from Table VII, ALEPH+DELPHI+L3 give no particular
indications. Indeed, their total $\chi^2=$27.9, 29.1, 28.6 and 31.2,
for the four pairs $(\alpha_s,m_h)$ considered in our analysis,
produce a maximum difference $\Delta\chi^2=+3.3$ so that even if the
top mass would be known with {\it infinite} precision to be 180 GeV
their data would be well consistent with {\it all} values of $m_h$
and $\alpha_s$.

Therefore, {\it if} we really want to explore the full potentiality
of LEP for a precise determination of $m_h$ (and $\alpha_s$) in the
standard electroweak theory, much more stringent tests have to be
performed. As stressed in \cite{zenro}, a precise scanning of the Z
resonance with 4 or more points at high statistics off peak cannot be
postponed anymore ( $\sim 90\%$ of the total events have been
collected at the pole ). Further, a high luminosity phase of LEP I,
where each Collaboration will detect millions of Z's per run and the
purely statistical errors will become negligible, is needed to obtain
a definitive consistency check of the systematics of the various
experiments.

\vskip 20pt
\centerline{ACKNOWLEDGEMENTS}
\par We thank Giampiero Passarino for many useful discussions.

\vskip 0.8cm

\newpage
\renewcommand{\arraystretch}{1.4}
\centerline{\large TABLES}

\vskip 2cm
\begin{center}
\begin{tabular}{lcccc}
\ ~~ &~~~ {ALEPH} &~~~ {DELPHI} &~~~ {L3} &~~~ {OPAL} \\ \hline\hline
$\Gamma_z(MeV)$ &~~~$2493.2\pm5.8$ & ~~~$2494.1\pm5.5$ &
\ ~$2504.0\pm5.3$ &~~~$2496.0\pm5.2$ \\ \hline
$\sigma_{had}$(nb) &~~~$41.62\pm0.10$ & ~~~$41.27\pm0.17$&
\ ~$41.41\pm0.11$&~~~$41.47\pm0.10$ \\ \hline
$R_e$ &~~~$20.63\pm0.13$ & ~~~$20.86\pm0.16$ &~~~$20.91\pm0.12$ &
\ ~$20.90\pm0.10$ \\ \hline
$R_{\mu}$ &~~~$20.95\pm0.12$ & ~~~$20.64\pm0.11$ &
\ ~$20.85\pm0.12$ &~~~$20.796\pm0.073$ \\ \hline
$R_{\tau}$ &~~~$20.68\pm0.12$ & ~~~$20.64\pm0.16$ &
\ ~$20.71\pm0.17$ &~~~$21.00\pm0.11$ \\ \hline
$A^{o}_{FB}(e)$ &~~~$0.0218\pm0.0055$ & ~~~$0.0221\pm0.0073$ &
\ ~$0.0125\pm0.0070$ &~~~$0.0081\pm0.0051$ \\ \hline
$A^{o}_{FB}(\mu)$ &~~~$0.0192\pm0.0039$ & ~~~$0.0168\pm0.0030$ &
\ ~$0.0164\pm0.0041$ &~~~$0.0137\pm0.0027$ \\ \hline
$A^{o}_{FB}(\tau)$ &~~~$0.0217\pm0.0044$ & ~~~$0.0210\pm0.0057$ &
\ ~$0.0305\pm0.0073$ &~~~$0.0183\pm0.0035$ \\ \hline
$A_e$ &~~~$0.129\pm0.017$ & ~~~$0.136\pm0.027$ &
\ ~$0.157\pm0.021$ &~~~$0.134\pm0.016$ \\ \hline
$A_{\tau}$ &~~~$0.136\pm0.015$ & ~~~$0.148\pm0.022$ &
\ ~$0.150\pm0.016$ &~~~$0.134\pm0.013$ \\ \hline
\end{tabular}
\end{center}

\vskip 1cm \noindent
{\bf Table I.}\ The experimental data from the four LEP
Collaborations.
\newpage
\vspace*{1.5cm}
\begin{center}
\begin{tabular}{lcccc}
\ $\alpha_s$ &~~~~ {$0.113$} &~~~~ {$0.125$} &~~~~ {$0.127$} &
\ ~~ {$0.130$} \\
\ $m_h$(GeV) &~~~~ {$100$} &~~~~ {$100$} &~~~~ {$500$} &
\ ~~ {$1000$} \\ \hline\hline
$\Gamma_z$(MeV) &~~~~$2495.8$ & ~~~~$2502.3$ &~~~~$2498.0$ &
\ ~~$2496.5$ \\ \hline
$\sigma_{had}$(nb) &~~~~$41.510$ & ~~~~$41.448$&~~~~$41.439$&
\ ~~$41.427$ \\ \hline
$R_e$ &~~~~$20.703$ & ~~~~$20.783$ &~~~~$20.785$ &~~~~$20.795$ \\
\hline
$R_{\mu}$ &~~~~$20.703$ & ~~~~$20.783$ &~~~~$20.785$ &~~~~$20.795$ \\
\hline
$R_{\tau}$ &~~~~$20.750$ & ~~~~$20.831$ &~~~~$20.833$ &
\ ~~$20.843$ \\ \hline
$A^{o}_{FB}(e)$ &~~~~$0.0174$ & ~~~~$0.0174$ &~~~~$0.0158$ &
\ ~~$0.0151$ \\ \hline
$A^{o}_{FB}(\mu)$ &~~~~$0.0174$ & ~~~~$0.0174$ &~~~~$0.0158$ &
\ ~~$0.0151$ \\ \hline
$A^{o}_{FB}(\tau)$ &~~~~$0.0174$ & ~~~~$0.0174$ &~~~~$0.0158$ &
\ ~~$0.0151$ \\ \hline
$A_e$ &~~~~$0.1524$ & ~~~~$0.1524$ &~~~~$0.1452$ &~~~~$0.1419$ \\
\hline
$A_{\tau}$ &~~~~$0.1524$ & ~~~~$0.1524$ &~~~~$0.1452$ &~~~~$0.1419$\\
\hline
\end{tabular}
\end{center}

\vskip 1cm \noindent
{\bf Table II.}\ We report the theoretical predictions at various
values of $\alpha_s(M_z)$ and $m_h$ for a fixed top-quark mass
$m_t=180$ GeV. These predictions have been obtained with the computer
code TOPAZ0 by G. Montagna, O. Nicrosini, G. Passarino,
F. Piccinini and R. Pittau.
\newpage
\vspace*{1.5cm}
\begin{center}
ALEPH

\vskip 30 pt
\begin{tabular}{lcccc}
\ $\alpha_s$ &~~~~ {$0.113$} &~~~~ {$0.125$} &~~~~ {$0.127$} &
\ ~~ {$0.130$} \\
\ $m_h$(GeV) &~~~~ {$100$} &~~~~ {$100$} &~~~~ {$500$} &
\ ~~ {$1000$} \\ \hline\hline
$\Gamma_z$ &~~~~$0.25$ & ~~~~$2.46$ &~~~~$0.68$ &~~~~$0.32$ \\ \hline
$\sigma_{had}$ &~~~~$1.21$ & ~~~~$2.96$&~~~~$3.28$&~~~~$3.72$ \\
\hline
$R_e$ &~~~~$0.32$ & ~~~~$1.38$ &~~~~$1.42$ &~~~~$1.61$ \\ \hline
$R_{\mu}$ &~~~~$4.24$ & ~~~~$1.94$ &~~~~$1.89$ &~~~~$1.67$ \\ \hline
$R_{\tau}$ &~~~~$0.34$ & ~~~~$1.58$ &~~~~$1.62$ &~~~~$1.84$ \\ \hline
$A^{o}_{FB}(e)$ &~~~~$0.64$ & ~~~~$0.64$ &~~~~$1.19$ &~~~~$1.48$ \\
\hline
$A^{o}_{FB}(\mu)$ &~~~~$0.21$ & ~~~~$0.21$ &~~~~$0.76$ &~~~~$1.10$ \\
\hline
$A^{o}_{FB}(\tau)$ &~~~~$0.96$ & ~~~~$0.96$ &~~~~$1.80$ &~~~~$2.25$\\
\hline
$A_e$ &~~~~$1.89$ & ~~~~$1.89$ &~~~~$0.91$ &~~~~$0.58$ \\ \hline
$A_{\tau}$ &~~~~$1.20$ & ~~~~$1.20$ &~~~~$0.38$ &~~~~$0.15$ \\
\hline\hline
TOTAL $\chi^2$ &~~~~$11.2$ & ~~~~$15.2$ &~~~~$13.9$ &~~~~$14.7$ \\
\hline
\end{tabular}
\end{center}

\vskip 1cm \noindent
{\bf Table III.}\ Individual and total $\chi^2$ from the ALEPH
Collaboration at various values of $\alpha_s(M_z)$ and $m_h$ for
$m_t=$180 GeV.
\newpage
\vspace*{1.5cm}
\begin{center}
DELPHI

\vskip 30 pt
\begin{tabular}{lcccc}
\ $\alpha_s$ &~~~~ {$0.113$} &~~~~ {$0.125$} &~~~~ {$0.127$} &
\ ~~ {$0.130$} \\
\ $m_h$(GeV) &~~~~ {$100$} &~~~~ {$100$} &~~~~ {$500$} &
\ ~~ {$1000$} \\ \hline\hline
$\Gamma_z$ &~~~~$0.10$ & ~~~~$2.22$ &~~~~$0.50$ &~~~~$0.19$ \\ \hline
$\sigma_{had}$ &~~~~$1.99$ & ~~~~$1.10$&~~~~$0.99$&~~~~$0.85$ \\
\hline
$R_e$ &~~~~$0.96$ & ~~~~$0.23$ &~~~~$0.22$ &~~~~$0.16$ \\ \hline
$R_{\mu}$ &~~~~$0.33$ & ~~~~$1.69$ &~~~~$1.74$ &~~~~$1.99$ \\ \hline
$R_{\tau}$ &~~~~$0.47$ & ~~~~$1.42$ &~~~~$1.45$ &~~~~$1.61$ \\ \hline
$A^{o}_{FB}(e)$ &~~~~$0.41$ & ~~~~$0.41$ &~~~~$0.74$ &~~~~$0.92$ \\
\hline
$A^{o}_{FB}(\mu)$ &~~~~$0.04$ & ~~~~$0.04$ &~~~~$0.11$ &~~~~$0.32$ \\
\hline
$A^{o}_{FB}(\tau)$ &~~~~$0.40$ & ~~~~$0.40$ &~~~~$0.83$ &~~~~$1.07$\\
\hline
$A_e$ &~~~~$0.37$ & ~~~~$0.37$ &~~~~$0.12$ &~~~~$0.05$ \\ \hline
$A_{\tau}$ &~~~~$0.04$ & ~~~~$0.04$ &~~~~$0.02$ &~~~~$0.08$ \\
\hline\hline
TOTAL $\chi^2$ &~~~~$5.1$ & ~~~~$7.9$ &~~~~$6.7$ &~~~~$7.2$ \\ \hline
\end{tabular}
\end{center}

\vskip 1cm \noindent
{\bf Table IV.}\ The same as in Table III for the DELPHI
Collaboration.
\newpage
\vspace*{1.5cm}
\begin{center}
L3

\vskip 30 pt
\begin{tabular}{lcccc}
\ $\alpha_s$ &~~~~ {$0.113$} &~~~~ {$0.125$} &~~~~ {$0.127$} &
\ ~~ {$0.130$} \\
\ $m_h$(GeV) &~~~~ {$100$} &~~~~ {$100$} &~~~~ {$500$} &
\ ~~ {$1000$} \\ \hline\hline
$\Gamma_z$ &~~~~$2.39$ & ~~~~$0.10$ &~~~~$1.28$ &~~~~$2.00$ \\ \hline
$\sigma_{had}$ &~~~~$0.83$ & ~~~~$0.12$&~~~~$0.07$&~~~~$0.02$ \\
\hline
$R_e$ &~~~~$2.98$ & ~~~~$1.12$ &~~~~$1.09$ &~~~~$0.92$ \\ \hline
$R_{\mu}$ &~~~~$1.50$ & ~~~~$0.31$ &~~~~$0.29$ &~~~~$0.21$ \\ \hline
$R_{\tau}$ &~~~~$0.06$ & ~~~~$0.51$ &~~~~$0.52$ &~~~~$0.61$ \\ \hline
$A^{o}_{FB}(e)$ &~~~~$0.49$ & ~~~~$0.49$ &~~~~$0.22$ &~~~~$0.14$ \\
\hline
$A^{o}_{FB}(\mu)$ &~~~~$0.06$ & ~~~~$0.06$ &~~~~$0.02$ &~~~~$0.10$ \\
\hline
$A^{o}_{FB}(\tau)$ &~~~~$3.22$ & ~~~~$3.22$ &~~~~$4.05$ &~~~~$4.45$\\
\hline
$A_e$ &~~~~$0.05$ & ~~~~$0.05$ &~~~~$0.32$ &~~~~$0.52$ \\ \hline
$A_{\tau}$ &~~~~$0.02$ & ~~~~$0.02$ &~~~~$0.09$ &~~~~$0.26$ \\
\hline\hline
TOTAL $\chi^2$ &~~~~$11.6$ & ~~~~$6.0$ &~~~~$8.0$ &~~~~$9.2$ \\
\hline
\end{tabular}
\end{center}

\vskip 1cm \noindent
{\bf Table V.}\ The same as in Table III for the L3 Collaboration.
\newpage
\vspace*{1.5cm}
\begin{center}
OPAL

\vskip 30 pt
\begin{tabular}{lcccc}
\ $\alpha_s$ &~~~~ {$0.113$} &~~~~ {$0.125$} &~~~~ {$0.127$} &
\ ~~ {$0.130$} \\
\ $m_h$(GeV) &~~~~ {$100$} &~~~~ {$100$} &~~~~ {$500$} &
\ ~~ {$1000$} \\ \hline\hline
$\Gamma_z$ &~~~~$0.00$ & ~~~~$1.47$ &~~~~$0.15$ &~~~~$0.01$ \\ \hline
$\sigma_{had}$ &~~~~$0.16$ & ~~~~$0.05$&~~~~$0.10$&~~~~$0.18$ \\
\hline
$R_e$ &~~~~$3.88$ & ~~~~$1.37$ &~~~~$1.32$ &~~~~$1.10$ \\ \hline
$R_{\mu}$ &~~~~$1.62$ & ~~~~$0.03$ &~~~~$0.02$ &~~~~$0.00$ \\ \hline
$R_{\tau}$ &~~~~$5.17$ & ~~~~$2.36$ &~~~~$2.30$ &~~~~$2.04$ \\ \hline
$A^{o}_{FB}(e)$ &~~~~$3.32$ & ~~~~$3.32$ &~~~~$2.28$ &~~~~$1.88$ \\
\hline
$A^{o}_{FB}(\mu)$ &~~~~$1.88$ & ~~~~$1.88$ &~~~~$0.60$ & ~~~$0.27$ \\
\hline
$A^{o}_{FB}(\tau)$ &~~~~$0.07$ & ~~~~$0.07$ &~~~~$0.51$ & ~~~$0.84$
\\ \hline
$A_e$ &~~~~$1.32$ & ~~~~$1.32$ &~~~~$0.49$ &~~~~$0.24$ \\ \hline
$A_{\tau}$ &~~~~$2.00$ & ~~~~$2.00$ &~~~~$0.74$ &~~~~$0.37$ \\
\hline\hline
TOTAL $\chi^2$ &~~~~$19.4$ & ~~~~$13.9$ &~~~~$8.5$ &~~~~$6.9$ \\
\hline
\end{tabular}
\end{center}

\vskip 1cm \noindent
{\bf Table VI.}\ The same as in Table III for the OPAL Collaboration.
\newpage
\vspace*{-1cm}
\begin{center}
ALEPH+DELPHI+L3+OPAL

\vskip 30 pt
\begin{tabular}{lcccc}
\ $\alpha_s$ &~~~~ {$0.113$} &~~~~ {$0.125$} &~~~~ {$0.127$} &
\ ~~ {$0.130$} \\
\ $m_h$(GeV) &~~~~ {$100$} &~~~~ {$100$} &~~~~ {$500$} &
\ ~~ {$1000$} \\ \hline
TOTAL $\chi^2$ &~~~~$47.3$ & ~~~~$43.0$ &~~~~$37.1$ &~~~~$38.1$ \\
\hline
\end{tabular}
\end{center}

\vskip 1cm \noindent
{\bf Table VII.}\ Total $\chi^2$ for the four Collaborations.

\vspace{2.5cm}
\begin{center}
\begin{tabular}{lcccc}
\ $\alpha_s$ &~~~~ {$0.113$} &~~~~ {$0.125$} &~~~~ {$0.127$} &
\ ~~ {$0.130$} \\
\ $m_h$(GeV) &~~~~ {$100$} &~~~~ {$100$} &~~~~ {$500$} &
\ ~~ {$1000$} \\ \hline\hline
ALEPH  &~~~~$10.2$  & ~~~~$14.3$ &~~~~$12.1$ &~~~~$12.5$ \\ \hline
DELPHI &~~~~$4.7$  & ~~~~$7.5$ &~~~~$5.9$ &~~~~$6.2$ \\ \hline
L3     &~~~~$8.4$ & ~~~~$2.8$ &~~~~$3.9$ &~~~~$4.8$ \\ \hline
OPAL   &~~~~$19.4$ & ~~~~$13.8$ &~~~~$8.0$ &~~~~$6.1$ \\ \hline\hline
TOTAL $\chi^2$ &~~~~$42.7$ & ~~~~$38.4$ &~~~~$29.9$ &~~~~$29.6$ \\
\hline
\end{tabular}
\end{center}

\vskip 1cm \noindent
{\bf Table VIII.}\ Total $\chi^2$ for the four Collaborations by
excluding the data for $A^{o}_{FB}(\tau)$.
\newpage
\vspace*{-1cm}
\begin{center}
ALEPH+DELPHI+L3+OPAL

\vskip 30 pt
\begin{tabular}{ccccc}
$\alpha_s$ &~~~~ {$0.113$} &~~~~ {$0.125$} &
\ ~~ {$0.127$} &~~~~ {$0.130$} \\
$m_h$(GeV) &~~~~ {$100$} &~~~~ {$100$} &~~~~ {$500$} &
\ ~~ {$1000$} \\ \hline\hline
$m_t$(GeV)=
170~~~ &~~~~$46.3$ & ~~~~$38.4$ &~~~~$38.3$ &~~~~$41.2$ \\ \hline
\phantom{$m_t$(GeV)}=
180~~~ &~~~~$47.3$ & ~~~~$43.0$ &~~~~$37.1$ &~~~~$38.1$ \\ \hline
\phantom{$m_t$(GeV)}=
190~~~ &~~~~$51.8$ & ~~~~$50.4$ &~~~~$38.9$ &~~~~$37.5$ \\ \hline
\end{tabular}
\end{center}

\vskip 1cm \noindent
{\bf Table IX.}\ Total $\chi^2$ for the four Collaborations at
various values of $m_t$.

\vspace{2cm}
\begin{center}
ALEPH+DELPHI+L3+OPAL

\vskip 30 pt
\begin{tabular}{ccccc}
$\alpha_s$ &~~~~ {$0.113$} &~~~~ {$0.125$} &
\ ~~ {$0.127$} &~~~~ {$0.130$} \\
$m_h$(GeV) &~~~~ {$100$} &~~~~ {$100$} &~~~~ {$500$} &
\ ~~ {$1000$} \\ \hline\hline
$m_t$(GeV)=
170~~~ &~~~~$40.7$ & ~~~~$32.8$ &~~~~$29.7$ &~~~~$31.2$ \\ \hline
\phantom{$m_t$(GeV)}=
180~~~ &~~~~$42.7$ & ~~~~$38.4$ &~~~~$29.9$ &~~~~$29.6$ \\ \hline
\phantom{$m_t$(GeV)}=
190~~~ &~~~~$50.1$ & ~~~~$46.6$ &~~~~$32.9$ &~~~~$30.3$ \\ \hline
\end{tabular}
\end{center}

\vskip 1cm \noindent
{\bf Table X.}\ Total $\chi^2$ for the four Collaborations at various
values of $m_t$ by excluding the data for $A^{o}_{FB}(\tau)$.
\end{document}